\documentclass[12pt]{article}

\usepackage{scicite}
\usepackage{graphicx}
\usepackage{hyperref}
\usepackage{caption}
\usepackage{times}

\newenvironment{sciabstract}{%
\begin{quote} \bf}
{\end{quote}}

\newcounter{lastnote}

\title{NeuroGen: activation optimized image synthesis for discovery neuroscience} 

\author
{Zijin Gu,$^{1}$ Keith Wakefield Jamison,$^{2}$ \\ Meenakshi Khosla,$^{1}$ Emily J. Allen,$^{4,5}$ Yihan Wu,$^{4}$ \\ Thomas Naselaris,$^{3}$ Kendrick Kay,$^{4}$ \\ Mert R. Sabuncu,$^{1}$ Amy Kuceyeski $^{2}\ast$\\
\\\textbf{}
\normalsize{$^{1}$Department of Electrical and Computer Engineering,}\\
\normalsize{Cornell University, Ithaca, New York, USA}\\
\normalsize{$^{2}$Department of Radiology,}\\
\normalsize{Weill Cornell Medicine, New York, New York, USA}\\
\normalsize{$^{3}$Department of Neuroscience,}\\
\normalsize{University of Minnesota, Minneapolis, Minnesota, USA}\\
\normalsize{$^{4}$Center for Magnetic Resonance Research(CMRR), Department of Radiology,}\\
\normalsize{University of Minnesota, Minneapolis, Minnesota, USA}\\
\normalsize{$^{5}$Department of Psychology,}\\
\normalsize{University of Minnesota, Minneapolis, Minnesota, USA}\\
\\
\normalsize{$^\ast$To whom correspondence should be addressed; E-mail:  amk2012@med.cornell.edu.}
}

\date{}

\begin{document} 


\baselineskip24pt


\maketitle


\begin{sciabstract}
  Functional MRI (fMRI) is a powerful technique that has allowed us to characterize visual cortex responses to stimuli, yet such experiments are by nature constructed based on a priori hypotheses, limited to the set of images presented to the individual while they are in the scanner, are subject to noise in the observed brain responses, and may vary widely across individuals. In this work, we propose a novel computational strategy, which we call NeuroGen, to overcome these limitations and develop a powerful tool for human vision neuroscience discovery. NeuroGen combines an fMRI-trained neural encoding model of human vision with a deep generative network to synthesize images predicted to achieve a target pattern of macro-scale brain activation. We demonstrate that the reduction of noise that the encoding model provides, coupled with the generative network's ability to produce images of high fidelity, results in a robust discovery architecture for visual neuroscience. By using only a small number of synthetic images created by NeuroGen, we demonstrate that we can detect and amplify differences in regional and individual human brain response patterns to visual stimuli. We then verify that these discoveries are reflected in the several thousand observed image responses measured with fMRI. We further demonstrate that NeuroGen can create synthetic images predicted to achieve regional response patterns not achievable by the best-matching natural images. The NeuroGen framework extends the utility of brain encoding models and opens up a new avenue for exploring, and possibly precisely controlling, the human visual system. 
\end{sciabstract}

\section*{Introduction}
Light rays reaching the retina are converted into bioelectrical signals and carried through the ophthalmic projections to the brain, where incoming signals are represented by corresponding neural activation patterns in the visual cortex \cite{WANDELL2007366}. The specific patterns of neural activation in response to visual stimuli are determined in part by the texture, color, orientation and content of the visual stimuli. The visual system has provided a rich model for understanding how brains receive, represent, process and interpret external stimuli, and has led to advances in understanding how the human brain experiences the world \cite{van1992information, thorpe1996speed}.

Much is known about how regions in the visual cortex activate in response to different image features or content. Our knowledge of stimulus-response maps has mostly been derived from identifying features that maximally activate various neurons or populations of neurons\cite{Hubel1962,Hubel1968}. Non-invasive techniques such as functional MRI (fMRI), are now one of the most utilized approaches for measuring human brain responses to visual (and other) stimuli \cite{Allen2021.02.22.432340, HCP}. The responses of early visual areas such as primary visual cortex (V1) have been studied using population receptive field (pRF) experiments wherein a participant fixates on a central dot while patterned stimuli continuously moved in the visual field \cite{WANDELL2007366}. Neurons in early visual areas have been found to be selective for stimulus location, but also other low-level stimulus properties such as orientation, direction of motion, spatial and temporal frequency \cite{hubel20208, deangelis1995receptive, de1980spatial, movshon1978spatial}. Recently, intermediate visual areas, like V2 or V4, 
were found to be responsive to textures, curved contours or shapes \cite{nandy2013fine, ziemba2016selectivity}. Late visual area activations have typically been explored by contrasting response patterns to images with varied content, e.g. to faces, bodies, text, and places. For example, the fusiform face area (FFA) \cite{kanwisher1997fusiform} involved in face perception, the extrastriate body area (EBA) \cite{downing2001cortical} involved in human body and body part perception, and the parahippocampal place area (PPA) \cite{epstein1998cortical} involved in perception of indoor and outdoor scenes, have been defined by contrasting patterns of brain activity evoked by images with different content. However, this approach has several limitations: the contrasts 1) are constructed based on a priori hypotheses about stimulus-response mappings, 2) are by nature limited to the set of images presented to the individual while they are in the scanner, 3) are subject to noise in the observed brain responses, and 4) may vary widely across individuals \cite{Seymour2018, Benson2018}. 

The recent explosion of machine learning literature has centered largely around Artificial Neural Networks (ANNs). These networks, originally inspired by how the human brain processes visual information \cite{Rosenblatt1958}, have proved remarkably useful for classification or regression problems of many types \cite{krizhevsky2012imagenet, simonyan2014very, he2016deep, toshev2014deeppose, belagiannis2015robust}. Common applications of ANNs are in the field of computer vision, including image segmentation \cite{girshick2014rich}, feature extraction \cite{krizhevsky2012imagenet, simonyan2014very} and object recognition \cite{sermanet2013overfeat}.
Meanwhile, in the field of neuroscience, researchers have incorporated ANNs into "encoding models" that predict neural responses to visual stimuli and, furthermore, have been shown to reflect structure and function of the visual processing pathway \cite{st2018feature,Khaligh-Razavi2014,Cichy2016,khosla2020cortical}.  Encoding models are an important tool in sensory neuroscience, as they can perform "offline" mapping of stimuli to brain responses, providing a computational stand-in for a human brain that also smooths measurement noise in the stimuli-response maps. ANNs' internal "representations" of visual stimuli have also been shown to mirror biological brain representations of the same stimuli, a finding replicated in early, mid and high-level visual regions \cite{Yamins2014,Cadena2019}. This observation has led to speculation that primate ventral visual stream may have evolved to be an optimal system for object recognition/detection in the same way that ANNs are identifying optimal computational architectures.

An alternative approach to understanding and interpreting neural activation patterns is decoding, in which the stimulus is reconstructed based on its corresponding neural activity response pattern. The presence of distinct semantic content in natural movies, e.g. object and action categories, has previously been decoded from fMRI responses with high accuracy using a hierarchical logistic regression graphical model \cite{10.3389/fnsys.2016.00081}. Beyond semantic content, natural scenes and even human faces can be reliably reconstructed from fMRI using generative adversarial network (GAN) approaches \cite{vanrullen2019reconstructing,mozafari2020reconstructing,st2018generative}. 
Encoding and decoding models, in conjunction with state-of-the-art generative networks, may also allow single neuron or neural population control. Recent work in macaques used an ANN-based model of visual encoding and closed-loop physiological experiments recording neurons to generate images specifically designed to achieve maximal activation in neurons of V4; the resulting synthetic images achieved higher firing rates beyond what was achieveable by natural images \cite{Bashivaneaav9436}. Moreover, by adopting a pretrained deep generative network and combining it with a genetic algorithm, realistic images were evolved to maximally activate target neurons in monkey's inferotemporal cortex \cite{PONCE2019999}. Both studies' results suggested the synthetic images uncovered some encoded information in the observed neurons that was consistent with previous literature, and, furthermore, that they evoked higher responses than any of the natural images presented. However, generative networks have not yet been applied to synthesize images designed to achieve maximal activation in macro-scale regions of the human visual cortex.

In this work, we build upon three recent advances in the literature. The first is the existence of the Natural Scenes Dataset (NSD), which consists of densely-sampled fMRI in eight individuals who each participated in 30-40 fMRI scanning sessions in which responses to 9,000-10,000 natural images were measured \cite{Allen2021.02.22.432340}. The second is in an interpretable and scalable encoding model, based on the NSD data, that performs accurate individual-level mapping from natural images to brain responses \cite{st2018feature}. The third is in the development of generative networks which are able to synthesize images with high fidelity and variety \cite{brock2018large,nguyen2016synthesizing}. Here, we propose a state-of-the-art generative framework, called NeuroGen, which allows synthesis of images that are optimized to achieve specific, predetermined brain activation responses in the human brain. We then apply this framework as a discovery architecture to amplify differences in regional and individual brain response patterns to visual stimuli. 

\section*{Results}
\subsection*{Encoding models accurately map images to brain responses and remove noise present in fMRI activation maps}
\begin{figure*}
\centering
\includegraphics[width=0.75\linewidth]{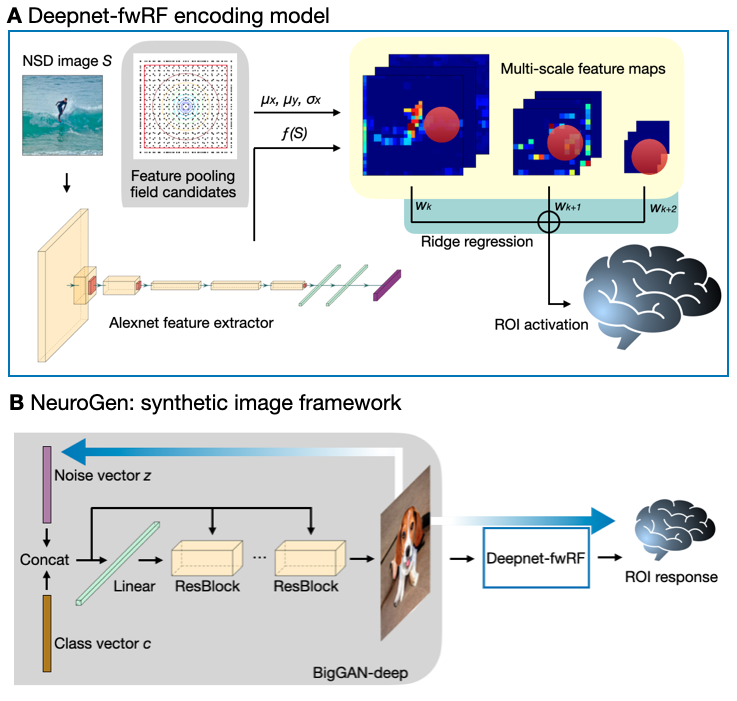}
\caption{Deepnet-fwRF encoding model and NeuroGen framework for synthetic image generation. \textbf{A} The deepnet-fwRF encoding model begins by passing an image through the Alexnet feature extractor, then applying a 2D Gaussian pooling receptive field to obtain multi-scale feature maps. Finally, ridge regression is applied to the multi-scale features to predict brain region-specific responses to the image. \textbf{B} The deepnet-fwRF encoding model is concatenated with a pretrained conditional generative network (BigGAN-deep) to synthesize images that are predicted to optimally match a desired response pattern (e.g. maximizing predicted activation in the fusiform face area). The optimized synthetic images are created in three steps. First, a single image for each of the 1000 classes in the conditional GAN is created from an initial truncated Gaussian noise vector; the resulting images are provided to the encoding model to obtain their predicted activation responses. Second, the 10 classes that give the predicted activation best matching the target activation are identified (e.g. those that give maximal predicted activation in the fusiform face area). Third, fine tuning of the noise vectors for each of the 10 synthetic images via gradient descent is performed; gradients flow from the encoding model's predicted response back to the synthetic image and to the noise vector that initializes the conditional GAN.}
\label{fig:1}
\end{figure*}

\begin{figure*}
\centering
\includegraphics[width=\linewidth]{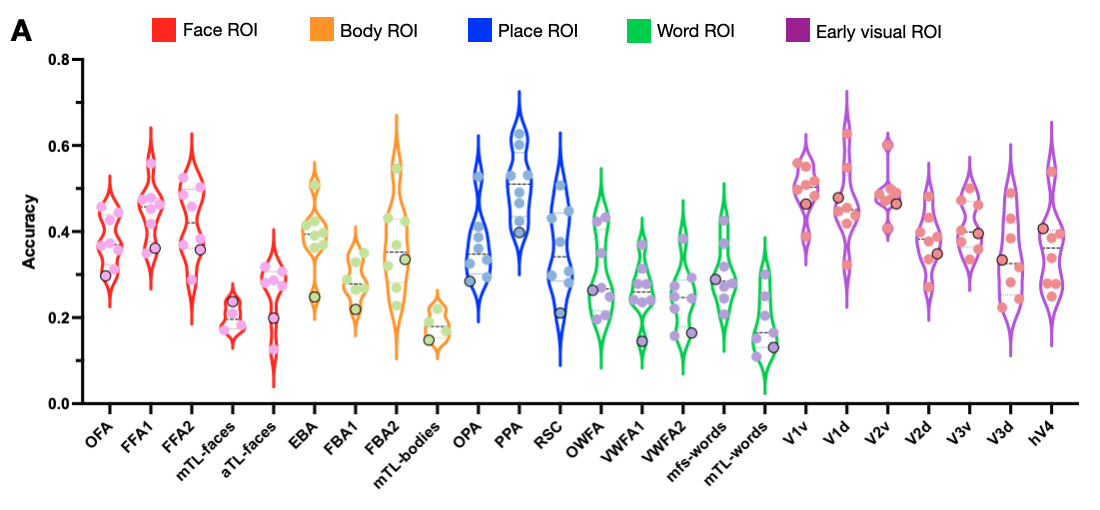}
\includegraphics[width=\linewidth]{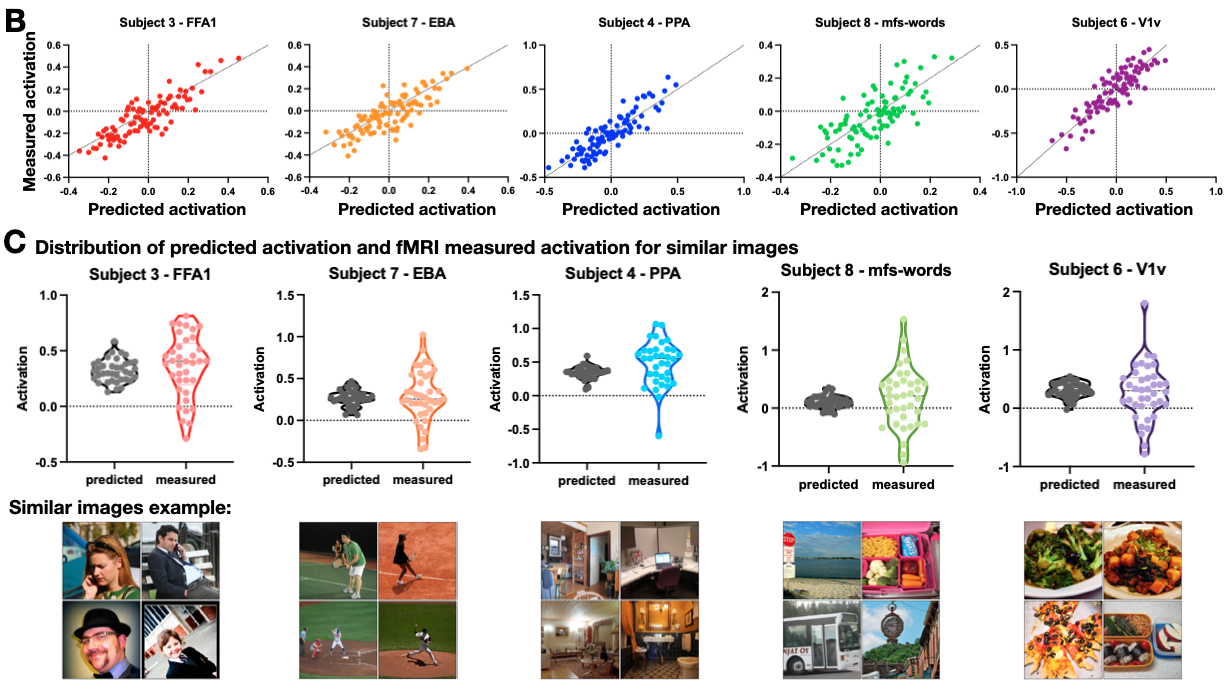}
\caption{The Deepnet-fwRF encoding model accurately predicts brain response to visual stimuli while smoothing noise present in fMRI-based individual image response maps. \textbf{A} Distribution of the individuals' hold-out prediction accuracies (Pearson correlation between predicted and observed for a hold-out test set of images) for all 24 brain regions, colored by category. Each point in the violin plot represents an individual; black outlines indicate the individuals whose results are shown in B) and in Figure \ref{fig:3}. \textbf{B} The scatter plot of the encoding model's predicted brain response vs observed brain response via fMRI for the shared set of 1000 test images for a given individual. Each point represents the mean activation of a sub-group of ten}
\label{fig:2}
\end{figure*}

First, subject and region-specific encoding models were built using the Deepnet-fwRF framework \cite{st2018feature} and the NSD data, which contained between 20k to 30k paired images and fMRI-based brain response patterns for each of eight individuals (Figure \ref{fig:1}A) \cite{Allen2021.02.22.432340}. Other than a shared set of 1000, the images were mutually exclusive across subjects. The encoding models first extract image features using AlexNet, a pre-trained deep neural network for image classification \cite{krizhevsky2012imagenet}. The encoding model then applies a 2D Gaussian pooling field to the image features to obtain a set of multi-scale feature maps. Ridge regression is used in the final step to predict individual regional activations from the multi-scale feature maps. For each of the eight subjects, we trained a separate model for each early/late visual region, of which there were at most 24. Regions were defined using functional localizer tasks \cite{Allen2021.02.22.432340} and some were missing in certain individuals. Figure \ref{fig:2}A shows the distribution of accuracies for each region's encoding model, colored by region category (face, body, place, word and early visual). The regional encoding model accuracy was assessed by calculating the Pearson correlation between each individual's model predictions with their measured fMRI regional activations for the shared set of approximately 1000 images (which were not used in the encoding model training). The accuracy for early, lower-order visual regions is generally higher than the accuracy in higher-order regions. Figure \ref{fig:2}B shows a scatter plot of the encoding model's predicted activation versus the observed fMRI measured activation for five example subject-ROI combinations. Each point represents the mean activation of a sub-group of ten images that are grouped together after rank ordering the thousand images from highest to lowest predicted activation. High consistency between the model predictions and actual fMRI measurements is observed, thus validating the reliability of the encoding models. 

\begin{figure*} [t!]
  \caption*{images that are grouped together after rank ordering the thousand images from highest to lowest predicted activation. \textbf{C} Distributions of the encoding models' predicted responses and fMRI measured responses for sets of similar images in the same 5 ROIs, with example images shown below). One region was visualized from each category: FFA1 (face), EBA (body), PPA (place), OWFA (word), V1v (early visual). Face ROIs: OFA - occipital face area; FFA - fusiform face area; mTLfaces - medial temporal lobe face area; aTLfaces - anterior temporal lobe face area. Body ROIs: EBA - extrastriate body area; FBA - fusiform body area;  mTLbodies - medial temporal lobe body area. Place ROIs: OPA - occipital place area; PPA - parahippocampal place area; RSC - retrosplenial cortex. Word ROIs: OWFA - occipital word form area; VWFA - visual word form area; mfswords - mid-fusiform sulcus word area; mTLwords - medial temporal lobe word area. Early visual ROIs: v - ventral; d - dorsal;}
\end{figure*}

To assess the amount of denoising the encoding model provides, we began by calculating each model's level of noise reduction by taking the ratio of the error in prediction to the error in measurement. To do this, we took the shared set of approximately 1000 images used for encoding model testing and selected those that were presented to the same subject 2-3 times (between 766-1000 images for each individual). For each of the images, we calculated the prediction error as the squared difference between the encoding model’s prediction and the mean of the 2-3 repeated measured fMRI responses and calculated the measurement error as variance of the 2-3 repeated measured fMRI responses. We took the ratio of these two numbers as the “rate of error reduction”. A value greater than 1 indicates that the prediction error is greater than the degree of noise in the measurement, whereas a value less than 1 can be interpreted as the prediction error being less than the noise level. The median error reduction over all 8 subjects' encoding models was 0.49 (0.43 to 0.55 IQR); all values were less than 1. This result indicates that the encoding model generally reduced measurement noise by around $50\%$, see Supplementary Figure S1 for a histogram of the rates of error reduction for all subjects' and all encoding models. In a second experiment to quantify encoding model noise reductions, we identified sets of images with similar content/features and compared the variability in the encoding model predictions and measured fMRI responses. We first filtered by label (“person”, “sports”, “indoor”, “text", and “food”) and then selected 30-40 that had subjectively similar content (see Figure 2C and Supplementary Figure S2-6). Figure \ref{fig:2}C shows the distributions of predicted response and fMRI measured response for the sets of similar images for the 5 subject-ROI combinations; the encoding model predictions had consistently smaller variance than the fMRI measurements with a range of 0.094-0.13 and 0.28-0.52, respectively. 

\subsection*{Optimized synthetic images achieve higher predicted responses than natural images}
\begin{figure*}
\centering
\includegraphics[width=0.48\linewidth]{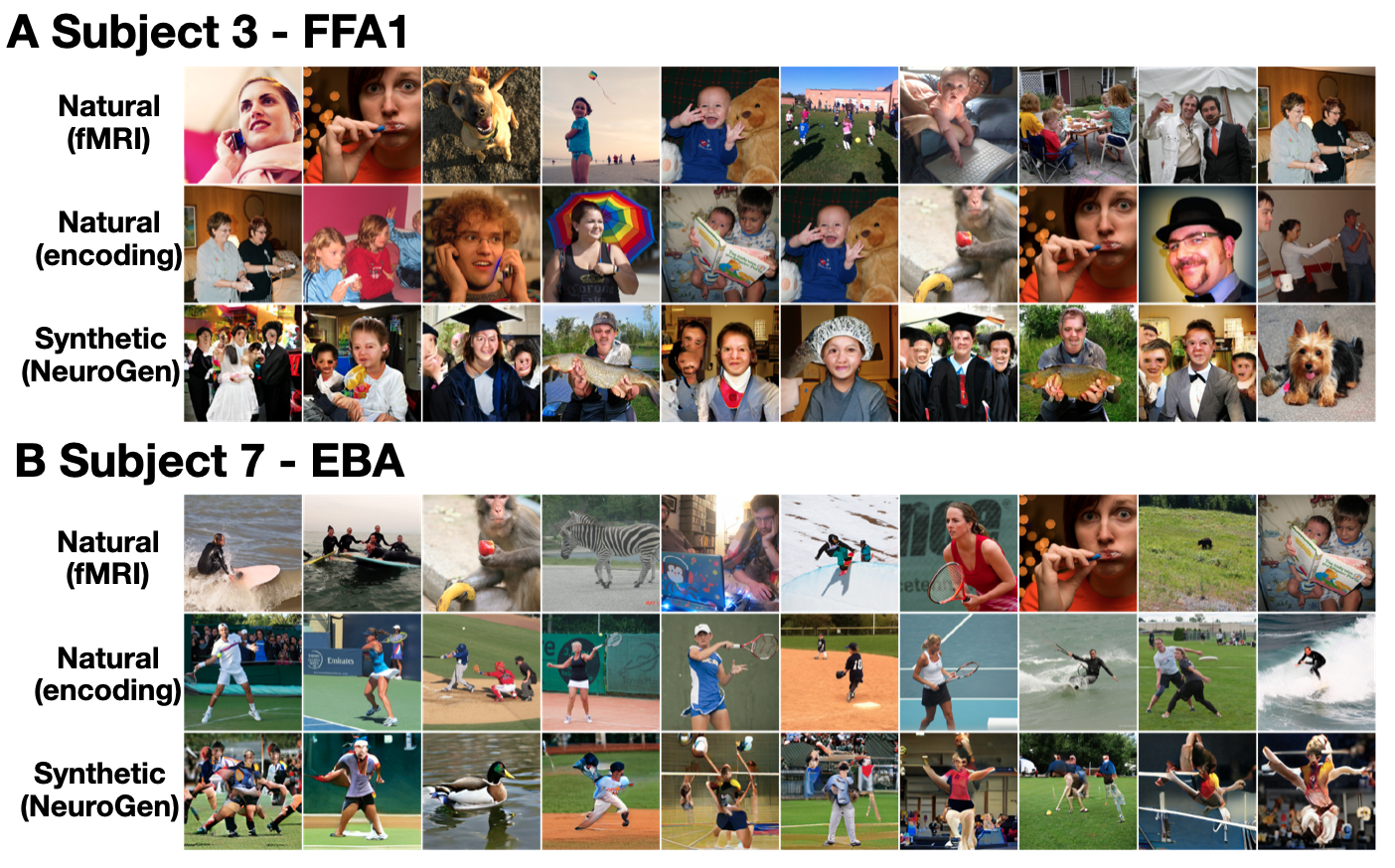}
\includegraphics[width=0.48\linewidth]{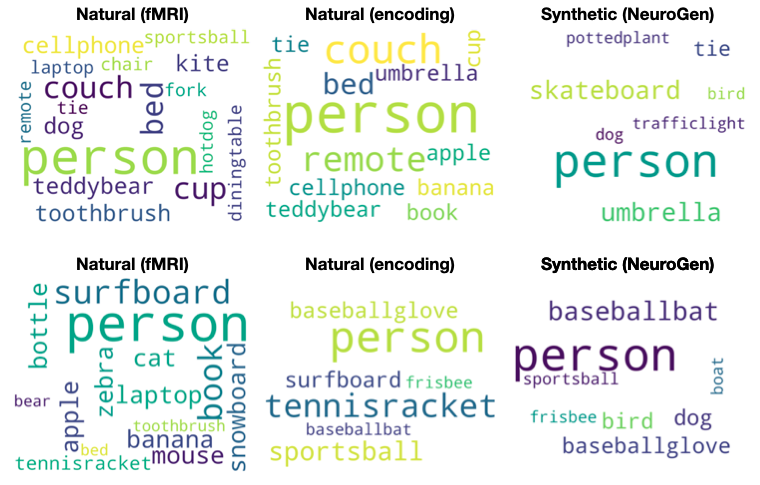}
\includegraphics[width=0.48\linewidth]{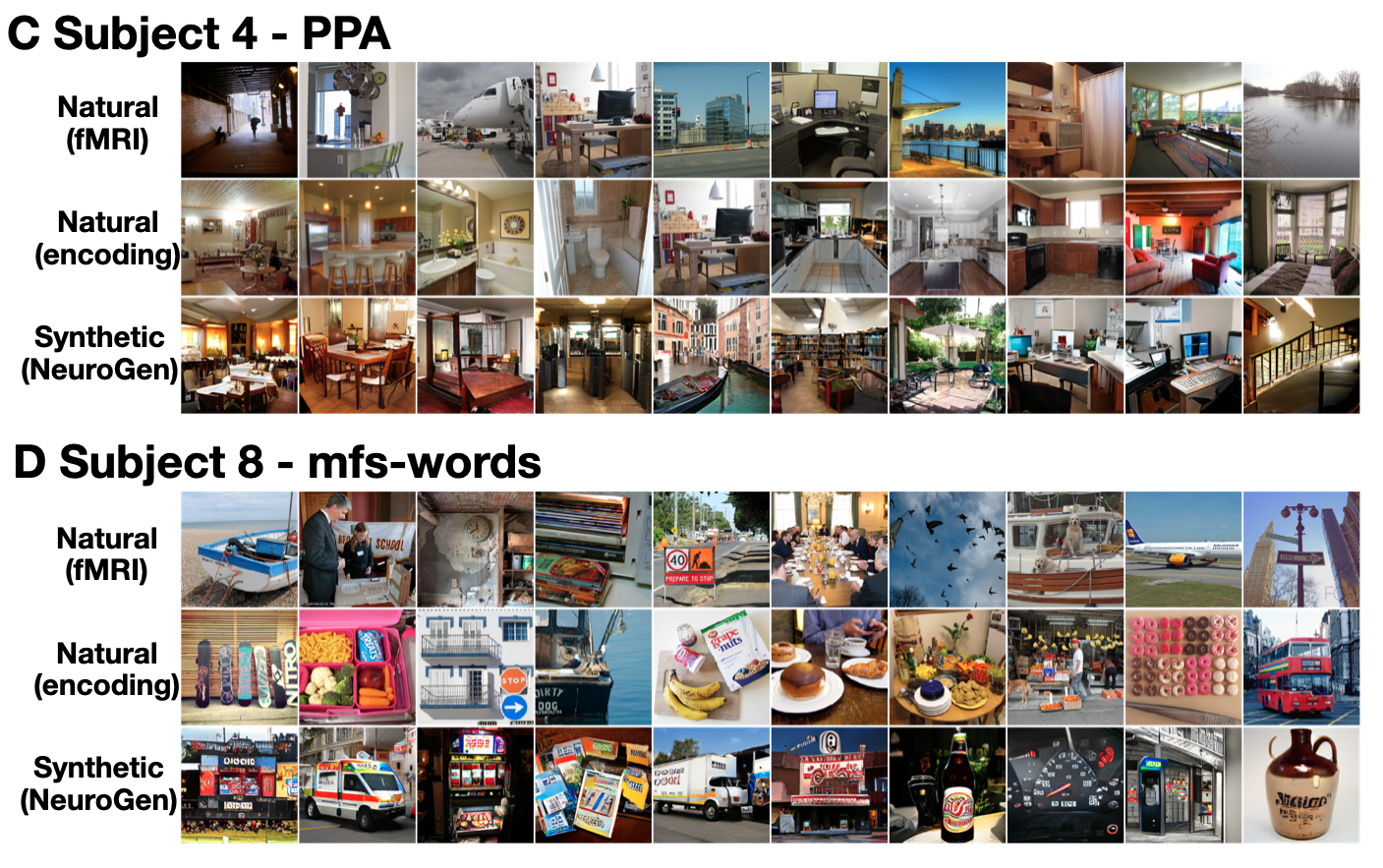}
\includegraphics[width=0.48\linewidth]{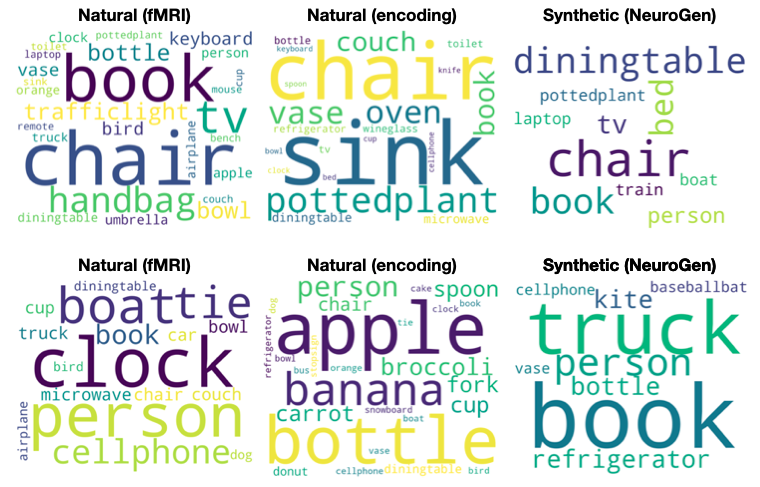}
\includegraphics[width=0.48\linewidth]{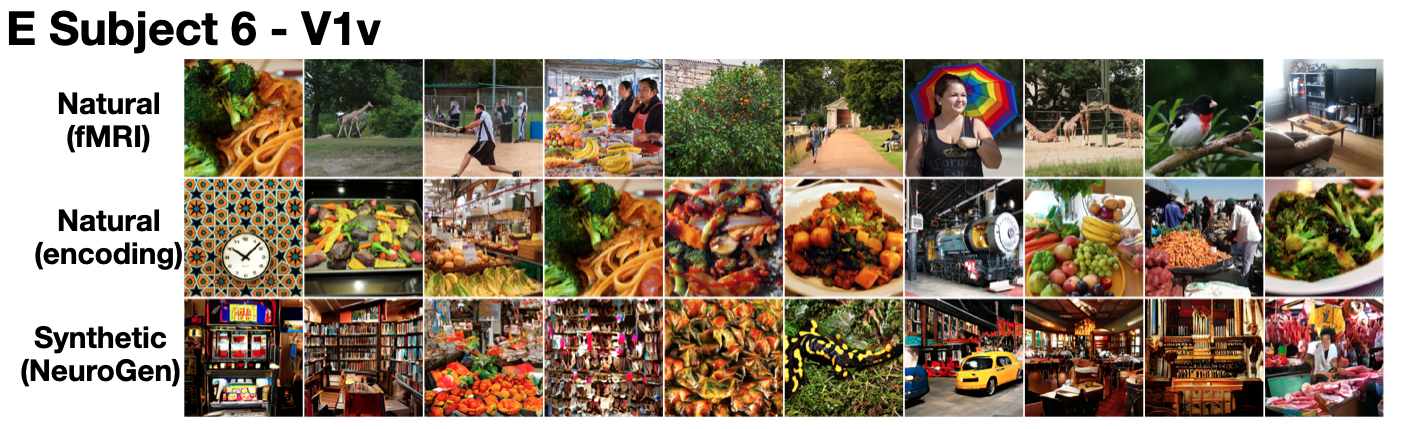}
\includegraphics[width=0.48\linewidth]{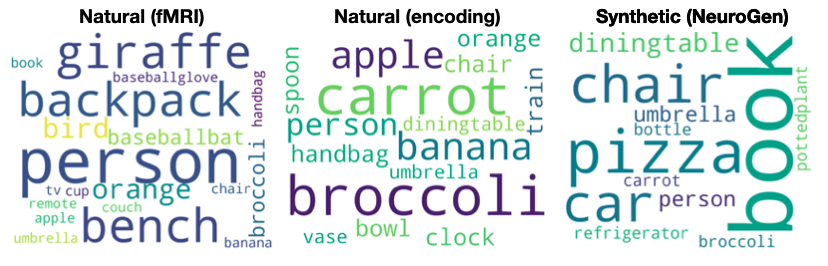}
\caption{Synthetic images generated to maximize predicted activation in five example visual regions in different individuals. In each panel, the top row contains the 10 natural images that had highest measured fMRI responses for that region, the second row contains the 10 natural images that had the highest encoding model predictions for that region, and the third row contains the 10 optimized synthetic images from NeuroGen for that region. Five example regions are shown, i.e. FFA1 (face), EBA (body), mfs-words (word), PPA (place) and V1v (early visual), each for a different individual that had median encoding model accuracy. The top encoding model images and NeuroGen images appear more consistent in their reflection of expected features/content compared to the top natural images with highest observed activation (e.g. fMRI top images include a giraffe for FFA1, a zebra for EBA and many lack text in the mfs-words area). The corresponding wordcloud plots show the image labels. \textbf{A} Subject 3's FFA1 region. \textbf{B} Subject 7's EBA region. \textbf{C} Subject 4's PPA region. \textbf{D} Subject 8's mfs-words region. \textbf{E} Subject 6's V1v region.}
\label{fig:3}
\end{figure*}

After validating the encoding model, we sought to generate images whose predicted activation optimally achieved certain criteria, e.g. had a maximal predicted activation in OFA. A conditional deep generative network (BigGAN-deep), based on the diverse set of natural images in ImageNet, was adopted as NeuroGen's synthetic image generator. By constructing the activation maximization scheme (Figure \ref{fig:1}B), which connects the encoding model and the image generator, gradients flowed from the regional predicted activation back to the noise vector to allow optimization. Because BigGan-deep is conditional, the class of the image is identified prior to synthesis which allows generation of more realistic synthetic images. We began optimization by first identifying the top 10 classes that gave the best match to our desired activation pattern. Once the top 10 classes were identified, the noise vector was then fine-tuned as previously described. We began by using this NeuroGen framework to generate images that give maximal predicted activation for a single region in turn. Figure \ref{fig:3}A, B, C, D, and E, shows the i) 10 natural images that had highest measured fMRI activation (top row) ii) 10 natural images that had the highest predicted activation from the encoding model (middle row) and iii) the 10 synthetic images optimized to achieve maximal predicted activation from the encoding model (bottom row) for a subject whose encoding model prediction accuracy was closest to the median accuracy. Word clouds representing the semantic content of all the top 10 images from the three sources are also provided by collecting the labels from the natural images and obtaining labels for the synthetic images using an automated image classifier \cite{ren2015faster}. The images and word clouds largely show alignment with expected image content and agreement across the three sources; however, when looking at the individual images, the content and features of both the natural and synthetic images with highest \textit{predicted} activations are more aligned with a priori expectations than the content and features of images with highest \textit{measured} activations. For example, the images with the highest observed activation include a giraffe for the face area, a zebra for the body area and many of the word area top images do not contain text. We posit that this is due to the encoding model predictions being less noisy than fMRI measurements. 

We do see a general agreement in the content/features of the synthetic images with what is expected from how the regions are defined. For FFA1 (face), the top 10 classes varied across individuals but were generally those that produced optimized images with human faces (ImageNet categories are terms like groom, mortarboard, shower cap, ice lolly, bow tie etc.) or dog faces (ImageNet categories were Pembroke, Basenji, Yorkshire terrier, etc.); the wordcloud reflects this as "person" is the dominant term. One interesting thing to note when comparing the natural and synthetic images for FFA1 is that the synthetic images tend to contain more than one person's face. Most of the top ImageNet classes for EBA (body) were sports-related (rugby ball, basket ball, volleyball, etc.) and the resulting images' most prominent features were active human bodies; the wordcloud again prominently features "person" but also sport-related terms. Typical indoor and outdoor place scenes emerged for place area PPA (wordcloud contained "dining table", "chair", "bed", etc.) and images with objects and/or places containing text were generated for word area mfs-words (wordcloud contained "book", "bottle" and "truck"). Synthetic images with high spatial frequency and color variety were produced when maximizing V1v (early visual), which generally responds to low-level image features such high frequency textures. In addition to the good agreement of the content and features of the optimized synthetic images with expectations, they are also predicted by the encoding model to achieve significantly higher activation than the top natural images (see Supplementary Figure S11). In fact, for all single region optimizations (over all 8 subjects and 24 regions), the synthetic images had significantly higher predicted activations than the top natural images. Supplemental Video S1 shows a video of the sets of top natural and synthetic images for all 8 individuals, for all 24 regions; Supplemental Figures S7-10 for wordclouds for all regions. Looking across all single region optimizations, there were some obvious differences in image content/features that emerged across individuals within the same region and within the same individual across regions in the same perception category, some of which we investigate further below. These results provide evidence that the NeuroGen framework can produce images that generally agree in content and features with \emph{a priori} knowledge of neural representations of visual stimuli, and may be able to amplify differences in response patterns across individuals or brain regions. 

\subsection*{Optimized synthetic images reflect and amplify features important in evoking individual-specific and region-specific brain responses}

\begin{figure*}
\centering
\includegraphics[width=0.48\linewidth]{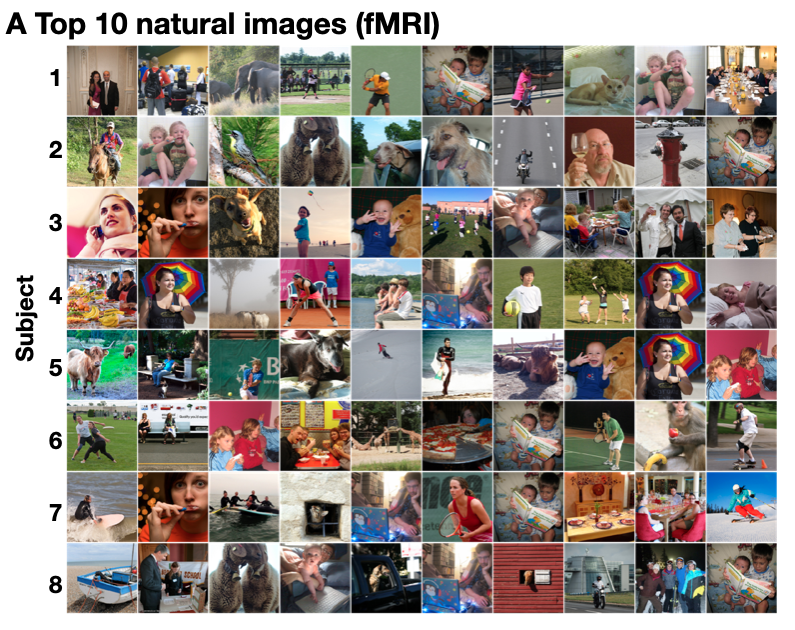}
\includegraphics[width=0.48\linewidth]{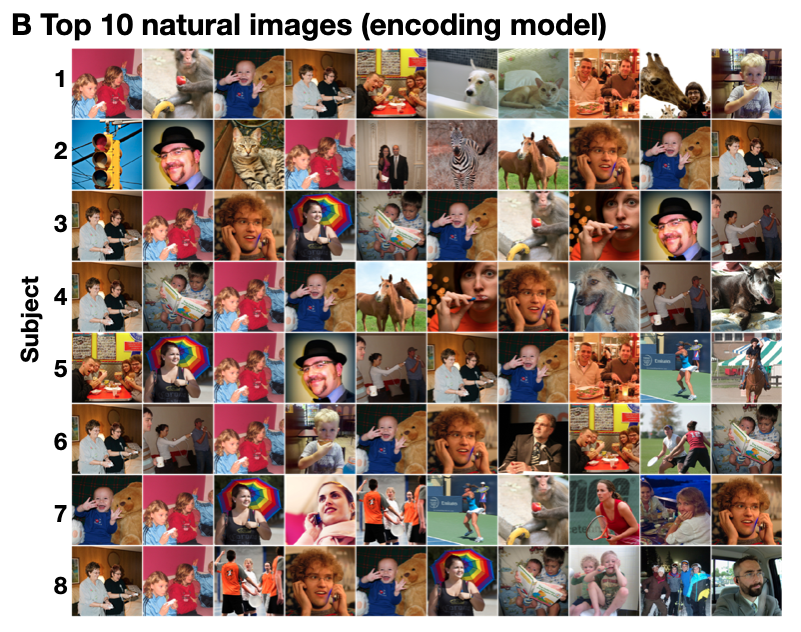}
\includegraphics[width=0.48\linewidth]{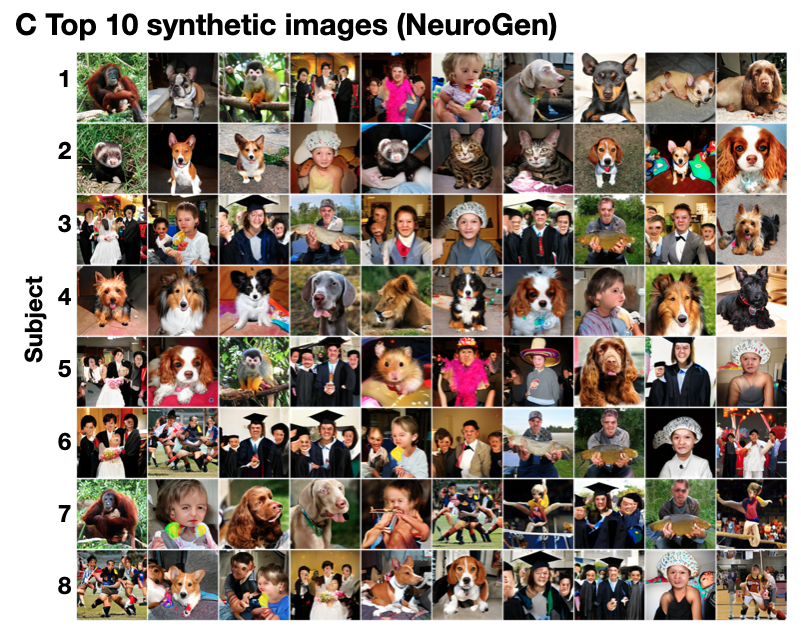}
\includegraphics[width=0.48\linewidth]{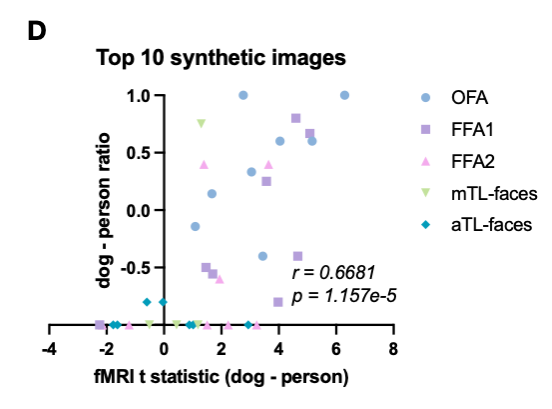}
\caption{Individual-specific and region-specific differences in face region responses are reflected in and amplified by the NeuroGen framework. \textbf{A}, \textbf{B} and \textbf{C} Sets of images that had the highest activation in FFA1 (fusiform face area 1) for all individuals, one per row, derived from three different sources. \textbf{A} Natural images that have the highest observed activation measured directly via fMRI. \textbf{B} Natural images that have the highest predicted activations from the encoding model. \textbf{C} Synthetic images that were created using NeuroGen. \textbf{D} The x-axis displays the dog vs. person preference from the observed fMRI data, quantified by the t-statistic of observed fMRI activations from all natural dog images compared to the observed activations from all natural people images, calculated for each of the five face areas in each of the eight individuals. The y-axes represent the dog vs person preference present in the top 10 synthetic images, calculated by taking the difference in the count of dog images minus the count of person images, divided by the total count of dog and person images. Values close to -1 indicate strong person preference and values close to 1 indicate strong dog preference. A significant correlation exists between the observed dog-person preference from the entire fMRI dataset and the dog-person preference in the top 10 synthetic images from NeuroGen (Spearman $r=0.6681$, $p=1.157e-5$).}
\label{fig:4}
\end{figure*}

\begin{figure*}
    \centering
    \includegraphics[width=\linewidth]{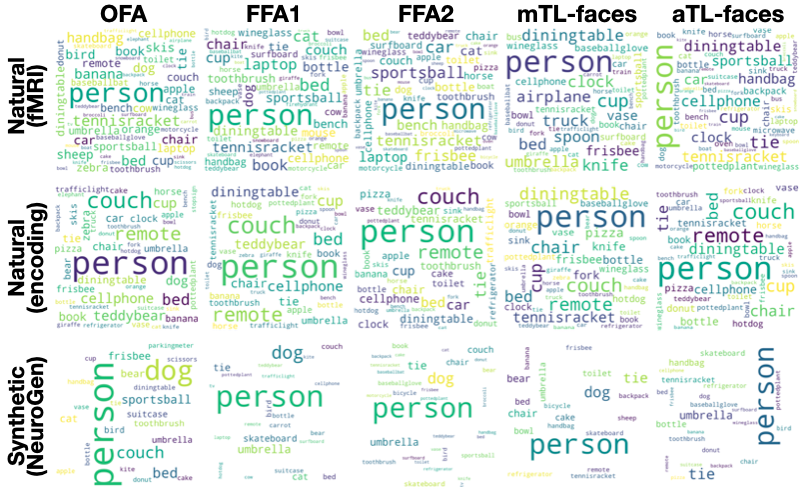}
    \caption{Face area preferences in semantic image content are reflected in the wordclouds illustrating image content for the top 10 natural and synthetic images. Each row represents the source of the top 10 images: natural images that have the highest observed activation measured directly via fMRI, natural images that have the highest predicted activations from the encoding model, and synthetic images that were created using NeuroGen. Each column represents one of the five face regions. The presence of the "dog" label (as well as, unsurprisingly, "person" label) can be appreciated most prominently in NeuroGen's synthetic images.}
    \label{fig:wordcloud}
\end{figure*}

One deviation in the content of the synthetic images from what was generally expected was the prevalence of dogs in all the five of face regions we modeled; 96 of the 350 top images (10 top images $\times$ 8 individuals $\times$ 5 face areas - 10 $\times$ 5 missing subjects' face areas) were of dog faces and 219 were of human faces. We observed some individuals' top 10 synthetic images all contained dogs while others had none. This imbalance also varied by brain region, one individual could have all synthetic images containing dogs for one face region while they would have fewer in another face region (see Supplemental Video S1). This apparent region- or individual-specific dog versus human preference was not apparent from the content of either the top 10 natural images giving highest observed or predicted activations. Figure \ref{fig:4}A, B and C, show all eight subjects' 10 natural images with the highest measured activation, 10 natural images with the highest predicted activation and 10 synthetic images, respectively, for an example face area (FFA1). For subject 3, nine out of ten synthetic images contain human faces, with only one dog. On the other hand, subject 4 had eight dog images, one human and one lion. Frequency based wordcloud plots of the top 10 synthetic images' labels for all individuals for each of the five face areas are shown in the third row of Figure \ref{fig:wordcloud}. The wordclouds confirm the regionally-varying prevalence of dogs in NeuroGen's synthetic images, which is not obvious from looking at either of the natural image wordclouds. We hypothesized that this individually and regionally varying dog/human preference in the synthetic images may be reflecting the actual underlying preferences in the data. To test this hypothesis, we calculated 1) the t-statistic of the measured fMRI activation from dog images (images in the NSD dataset that had a single label "dog") and the measured fMRI activation from images of humans (images in the NSD dataset that had a label "person" and any of the following: "accessory", "sports")  and 2) the top 10 and top 100 synthetic images' dog vs human image ratios, calculated as ((number of dog images - number of person images)/(number of dog images + number of person images)). We calculated these two measures (dog vs. person t-stat and synthetic image ratio) for all eight individuals and all five face ROIs (see Figure \ref{fig:4}D and E). We found a significant correlation between the t statistic and image ratios for NeuroGen's top 10 synthetic images (Spearman rank $r=0.6681$, $p=1.157e-5$); this correlation was even higher when considering NeuroGen's top 100 synthetic images (Spearman rank $r=0.7513$, $p=1.987e-7$), see Supplementary Figure S12. Correlations with the top 10 natural images from the encoding model and fMRI measurements were also significant, but not as strong as the top 10 synthetic images correlation (Spearman rank $r=0.4468$, $p=7.126e-3$ and $r=0.2948$, $p=0.086$), respectively. These results show that previously identified "face" regions in the human visual cortex also respond robustly to dog faces, and, furthermore, that the dog/human balance in response patterns varies across individuals and brain regions. These results highlight NeuroGen's potential as a discovery architecture, which can be used to amplify and concisely summarize (even with only 10 images) region-specific and individual-specific differences in neural representations of visual stimuli. 

\begin{figure*}
\centering
\includegraphics[width=0.48\linewidth]{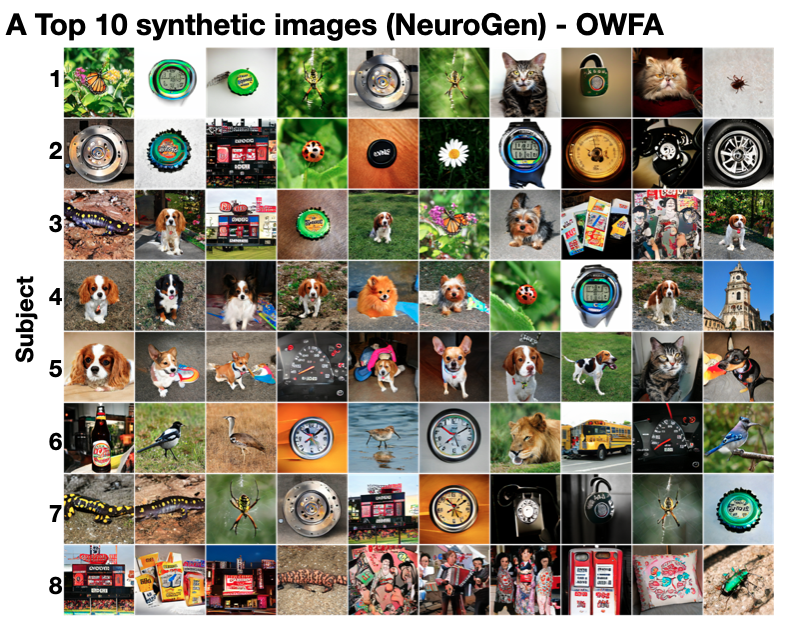}
\includegraphics[width=0.48\linewidth]{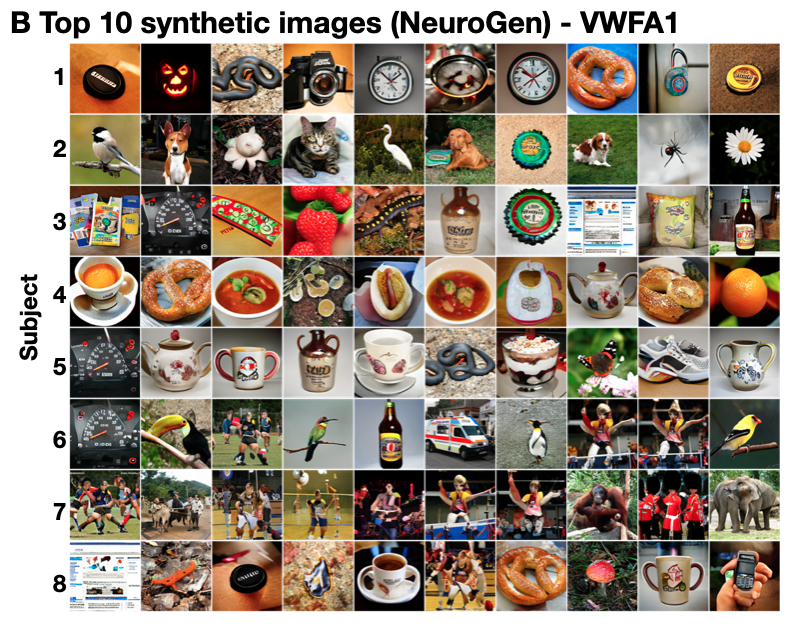}
\includegraphics[width=0.48\linewidth]{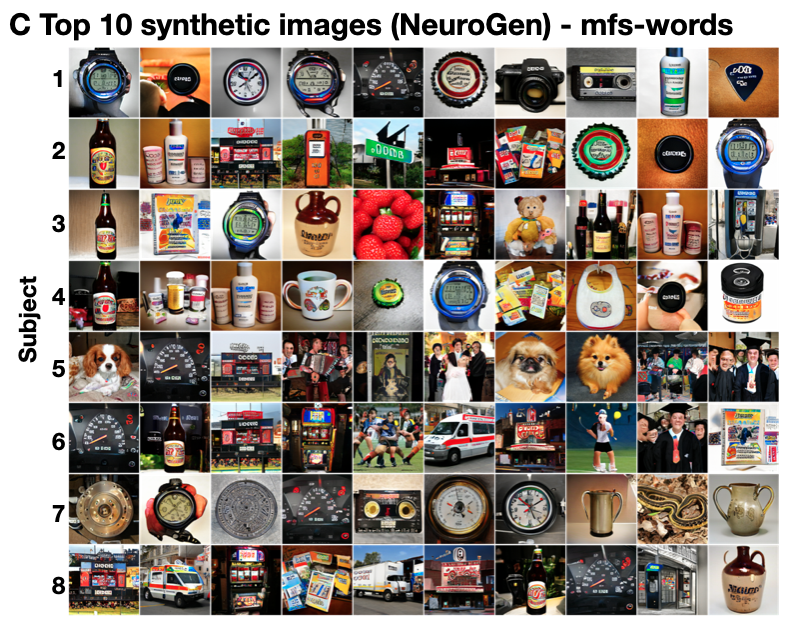}
\includegraphics[width=0.48\linewidth]{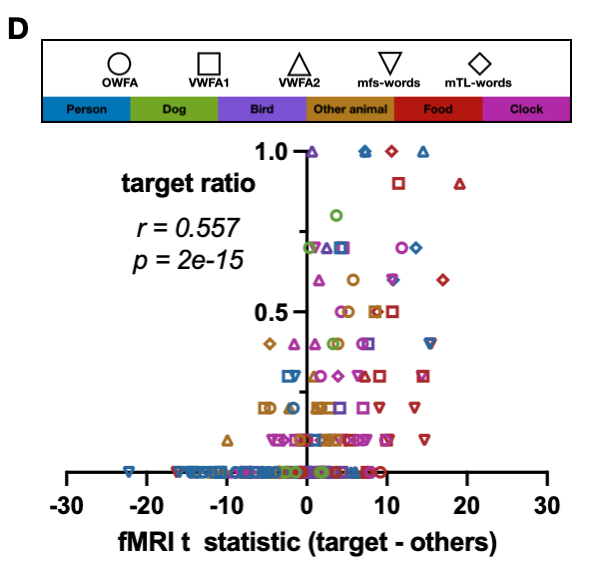}
\caption{Individual-specific and region-specific differences in word area responses are reflected in and amplified by only 10 images from the NeuroGen framework. Sets of images that had the highest activation in \textbf{A} OWFA (occipital word form area), \textbf{B} ventral word form area 1 (VWFA1) and \textbf{C} mfs-words, for all eight individuals (one per row). \textbf{D} The x-axis displays the activation contrast for six categories of natural images (human, dog, bird, other animal, food and clock) from the observed fMRI data, quantified by the t-statistic of observed fMRI activations from natural images of the category in question compared to the observed activations from all other natural images (not containing that item). The y-axis represents the proportion of top 10 synthetic images that contain that item. A significant correlation exists between the observed image category contrasts from the entire fMRI dataset and the proportions of that image category in the top 10 synthetic images from NeuroGen (Spearman rank $r=0.557$, $p=2e-15$).}
\label{fig:5}
\end{figure*}

Another unexpected observation of individual and regional variability was found in the word areas. Because the images in NSD are natural, they do not solely contain text; therefore many of the top natural and synthetic images for the word-preferred regions contained objects or scenes with integrated text (scoreboard, packet, cinema, bottle cap, sign, etc.), see Figure \ref{fig:5}. However, images with integrated text were only a portion of the word form area top images; many also contained humans, dogs, cats, birds, other animals, food and clocks. We aimed to test if the content of the synthetic images from NeuroGen could accurately reflect underlying patterns in the measured activation data, even when considering several categories of images. Therefore, we calculated the proportion of each individual's top 10 images that contained objects in each of the six categories of interest (humans, dogs, cats, birds, other animals, food and clocks) for each of the top 5 word areas. These six categories of interest were selected based on the most common categories in the top 10 synthetic images across all 5 word form areas. We then correlated each individual's synthetic image proportions with their t-statistic of measured activation for images containing that target object (and not any objects in the other 5 categories) contrasted against the measured activation for all images not containing that object. Categories were included for a given word form area's analysis if there existed at least 10 images of that category out of the 240 total top natural or synthetic images for that region (8 individuals x 10 images x 3 sources). Figure \ref{fig:5}D shows the scatter plot of the image proportions versus the t-statistic giving the activation contrast for synthetic images of a given category (indicated by color) for the various word form regions (indicated by shape), which were significantly correlated (Spearman rank $r=0.557$, $p=2e-15$). The proportion of the 10 natural images with highest encoding model predicted activation and highest measured activation via fMRI had a somewhat weaker but still significant correlation (Spearman rank $r=0.4545$, $p=3.781e-10$ and Spearman rank $r=0.1897$, $p=0.0127$, respectively). These results again highlight the utility of NeuroGen as a discovery architecture that can concisely uncover, even across several categories of images at once, neural representation variability across individuals and brain regions within an individual.

\subsection*{Optimized synthetic images have more extreme predicted co-activation of region-pairs than natural images}

\begin{figure*}
\centering
\includegraphics[width=\linewidth]{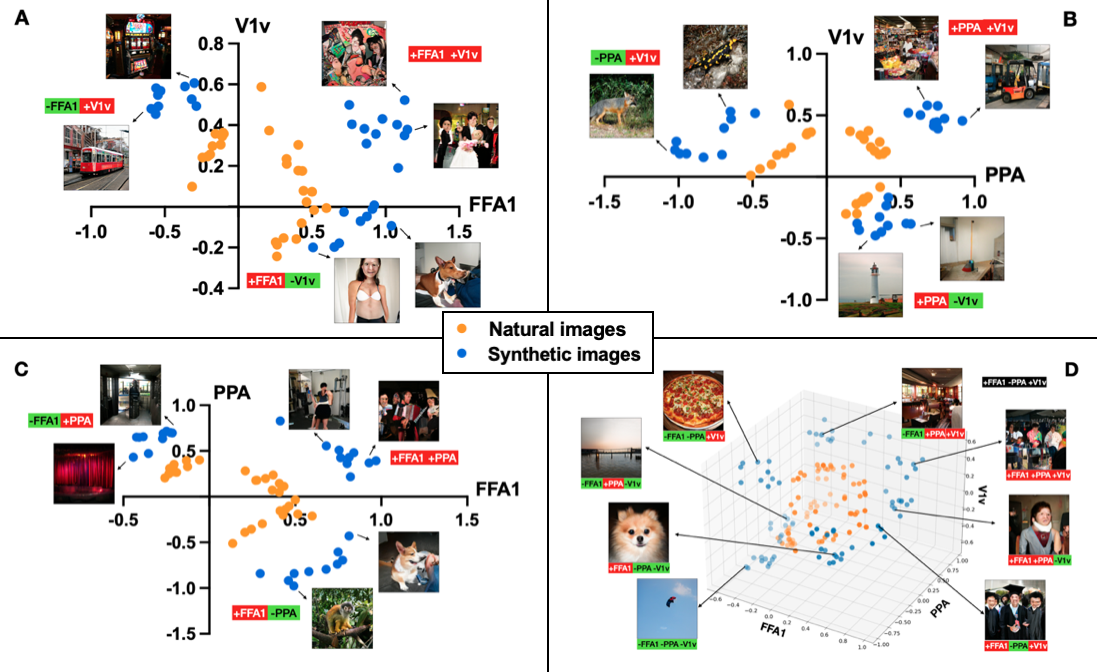}
\caption{Multiple-region optimization creates synthetic images with generally more extreme predicted activation when compared to natural images. Scatter plots show the predicted response from top 10 natural and synthetic images that either jointly maximize the sum of the two regions (+ROI1+ROI2) or jointly maximize one and minimize the other (+ROI1-ROI2) and (-ROI1+ROI2), for the following region-pairs \textbf{A} FFA1 and V1v, \textbf{B} PPA and V1v and \textbf{C} FFA1 and PPA. \textbf{D} shows all sets of three-region optimization combinations for FFA1, PPA and V1v together. Typical synthetic image examples are shown for each dual and triple optimization; the optimization specifics are listed below each synthetic image (green = that region's minimization, red =  that region's maximization).}
\label{fig:6}
\end{figure*}

The NeuroGen framework is flexible and can be used to synthesize images predicted to achieve an arbitrary target activation level for any or all of the 24 regions for which we have encoding models. We now provide examples of two-region and three-region optimization; specifically, we aimed to create synthetic images that jointly maximize and/or minimize predicted activation in the target regions together. For example, two region optimization could either be joint maximization (+ROI1+ROI2), or joint maximization of one region and minimization of the other, or maximizing (+ROI1-ROI2 or -ROI1+ROI2). Figure \ref{fig:6} shows the results of three example two region optimizations for subject 8's A) FFA1 (face) and V1v (early visual) regions, B) PPA (place) and V1v (early visual) regions, C) FFA1 (face) and PPA (place) regions, and one example of three region optimization (FFA1, PPA and V1v). When jointly maximizing FFA1 or PPA with V1v, NeuroGen's synthetic images are of places or faces with an abundance of texture (e.g. comic book, steel drum and kimono labels for FFA1 and toy shop, bookstore, slot labels for PPA). Alternatively, when jointly maximizing FFA1 or PPA and minimizing V1v, we see human/dog faces or places/indoor scenes with flat, non-textured colors in both the foreground and background (e.g. neck brace, ice lolly and bikini for FFA1 and beacon, pier, container ship labels for PPA). Natural images show a similar pattern, although they are not as obvious and appear less consistent (see Supplementary Figure S13). In addition, we see that the top 10 optimized synthetic images generally had significantly more extreme predicted activation values in the desired direction than the top 10 natural images for most cases (see Figure \ref{fig:6} and Supplementary Figure S5). Both the two- and three- region optimizations demonstrate the synthetic images push the predicted activations past the boundaries of the natural image activations; there is a clear separation of the orange and blue points representing these values. These results demonstrate that the NeuroGen framework can be applied to create synthetic images predicted to achieve optimized activation in more than one region at the same time; the example synthetic images provided align with expectations from prior knowledge. Furthermore, they suggest that the benefit of NeuroGen's synthetic images over natural ones in terms of pushing the predicted activation levels to those not achievable by natural images.

\section*{Discussion}
Here, we introduce NeuroGen, a novel architecture designed to synthesize realistic images predicted to maximize or minimize activation in pre-selected regions of the human visual cortex. NeuroGen leverages three recent scientific advances: 1) the development of encoding models that can accurately predict brain responses to visual stimuli \cite{st2018feature}, 2) in deep generative networks' abilities to synthesize high-fidelity and variety images \cite{brock2018large} and 3) in the recent curation of the Natural Scenes Dataset (NSD), which consists of tens of thousands of paired images and human brain responses \cite{Allen2021.02.22.432340}. We showed that encoding models trained on the NSD data could accurately maps images to their neural representations in individual subjects, and, importantly, that the encoding model smooths noisy measured fMRI response maps. Once the encoding model was validated, we used the NeuroGen framework to concisely amplify and reveal individual and regional preferences for certain image types. We began by using NeuroGen to create synthetic images that were predicted to maximize a single region's activation response. The resulting synthetic images agreed with expectations from previous knowledge of regional neural representations of visual stimuli and, furthermore, the predicted activations from synthetic images were significantly higher than from natural images. Once NeuroGen was validated, it was used as a discovery architecture to uncover region-specific and individual-specific visual cortex response patterns. Our main discovery was a remarkable, previously not well-described balance of dog-human preferences in face areas that both varied across face regions and individuals. The synthetic image human/dog preference ratios were validated by showing strong, significant correlations with dog/human preference ratios calculated by contrasting measured fMRI activations in response to thousands of dog and human images. Secondly, we used the NeuroGen framework to show that the content of images preferentially activating word form areas were of a wide variety, including humans, dogs, cats, birds, other animals, clocks, food and more. Despite the fact that several categories of images were represented in the word form areas, we again validated the top 10 synthetic image ratios by showing significant correlations with underlying preferences extracted by contrasting measured fMRI activations in response to hundreds of images. Finally, we extended the single region analysis to demonstrate the capacity of the NeuroGen framework in optimizing activation for two or three regions at a time. We found that these two- and three-region optimizations not only produced images that agreed with expectations, but also provided significantly more extreme predicted activations than natural images, above and beyond the activation levels observed in response to the best-matching natural images. Taken together, these results validate and demonstrate that the NeuroGen framework can create new hypotheses for neuroscience and thus facilitate a tight loop between modeling and experiments, and thus is a robust and flexible discovery architecture for vision neuroscience.

The visual system provides an excellent model with which to understand how organisms experience the environment. Mapping the visual system's neural representations of external stimuli has often centered around identifying features that maximally activate various neurons or populations of neurons\cite{Hubel1962,Hubel1968}. This "activation maximization" approach, more commonly called the tuning curve approach, has lead to discoveries of visual regions that selectively respond to specific patterns \cite{Kobatake1994,WANDELL2007366} or images with a certain content, most prominently, faces \cite{kanwisher1997fusiform,Tsao2006}, places \cite{epstein1998cortical}, bodies \cite{downing2001cortical,Popivanov2014} and visual words \cite{Baker2007}. This "activation maximization" approach using \emph{in vivo} measurements is by nature limited to the stimuli presented while observing responses, which is in turn biased by \emph{a priori} hypotheses. There may be more complex, obscure stimuli-response maps that exist (for example, perhaps, to images of a dog riding a bicycle) but are not tested due to our limited imaginations or the lack of representation of that type of image in natural image sets. In addition, fMRI can be very noisy, and response maps to a handful of images (or even hundreds of them) are quite noisy even within an individual, let alone across the population. Encoding models that can perform "offline" mapping of stimuli to brain responses can provide a computational stand-in for a human brain that also smooths measurement noise in the stimuli-response maps. NeuroGen's framework that couples the benefits of encoding models with recent advances in generative networks in creating naturalistic-looking images, may prove to be an advance in discovery neuroscience that is more than the sum of its parts.

Only a few previous studies have used generative networks to create synthetic images made to achieve activation maximization of single neurons or populations of neurons in non-human primates. Both used closed-loop physiological experimental designs to record and optimize neuronal responses, e.g. maximize firing rates, to synthetic images \cite{PONCE2019999, Bashivaneaav9436}. One synthesized images by directly optimizing in image space using an ANN model for the brain's ventral visual stream \cite{Bashivaneaav9436}, and the other synthesized images in code space via a genetic algorithm to maximize neuronal firing in real time \cite{PONCE2019999}. Both of these studies successfully demonstrated that single neurons or neuronal populations in monkeys can be controlled via optimization of synthetic images using generative networks. Until now, no work has attempted to create a generative framework that synthesizes images to maximize activation in macro-scale regions of the human brain. One difference in our framework, other than the species in question, is the use of a conditional generator network that requires the identification of an image class before synthesis. We wanted our framework to synthesize images that were as natural-looking as possible for two reasons: because our encoding model was trained on natural images and because future work will include presentation of these synthetic images to humans while they are undergoing fMRI to test if they achieve activation above and beyond the best natural images.

Our main discovery using the NeuroGen architecture was a previously not well-described balance of dog-person preference in face areas, which varied over individuals within the population and regions within an individual. After inspecting the content of the top 10 images from NeuroGen, we noted an abundance of dog faces in addition to human faces that was not obvious in the top 10 natural images with the highest measured or predicted activation; this was also apparent from looking at the face regions' wordclouds. We showed that the dog-human preference ratio observed in NeuroGen's synthetic images was reflected in the underlying data by observing a strong, significant correlation with the t-statistic of the measured activation (via fMRI response) from dog images versus human images. One idiosyncrasy of the ImageNet data used to train our generative network is its prevalence of dogs; 120 out of the approximately 1000 ImageNet classes are dog breeds and, furthermore, dog images in ImageNet generally feature close-ups of dog's faces. This over-representation of dogs in the ImageNet database could have biased NeuroGen to more easily identify and amplify any existing dog-human preferences in the underlying data. In addition, measured contrasts for some regions in some individuals showed a clear preference for dog faces over human faces (t-statistic $>4$), which we conjecture could be due to either differences in visual attention between the two categories or the fact that the NSD "person" images used to calculate the contrast are not all close-ups of human faces while the dog images do tend to be close up dog faces. One of the few studies comparing humans' neural representations of dog faces and human faces showed very similar response maps to both species, with lingual/medial fusiform gyri being the only region having higher activations for dog over human faces \cite{Blonder2004}. We conjecture that differences in our findings may be due to their population-level approach to identifying differences in neural representations, as they used coregistered contrast maps to identify group-level, voxel-wise significance. Other studies have shown that human face areas also respond to mammals, although again at a population level the activation in response to mammals was not stronger than responses to humans \cite{Downing2006}. We see that NeuroGen's dog-human balance in response patterns varies widely over individuals and brain regions, indicating that population-level approaches may not be adequate for creating stimulus-response maps.

While humans' neural representations of faces, places and bodies are generally robust across the population, it has been shown that word form responses can vary based on an individual's experience \cite{Baker2007,Kanwisher2010}. Our findings generally revealed more divergence in the word form area preferred content across individuals than other categories of visual regions. This large individual-level variability in preferred image content, including images of several very different categories (humans, dogs, cats, birds, food, clocks), could be due to the effect of individual experience in forming the neural representations in these word form areas. On the other hand, these areas do tend to be quite small and more susceptible to noise in the measured activation patterns which were used to define the regions leading to more population-level divergence \cite{brett2002problem}. The natural images in the NSD dataset used to create the encoding model also did not contain isolated text, which could further contribute to noise in applying the NeuroGen framework to word form areas. However, many of the synthetic images were derived from categories that contain items with text, including "odometer", "comic book", "book jacket", "street sign", "scoreboard", "packet" and "pill bottle". The word form regions did also overlap regions in other categories (see Supplementary Figure S14-22), including face and place areas. This overlap could explain the presence of dogs and humans (and possibly other mammals) but it does not explain, for example, the strong presence of food images in many of the individuals' top images (see Figure \ref{fig:4}B). Despite these potential shortcomings, using only 10 images, NeuroGen was able to reflect measured, underlying preferences across several categories for these complex and widely varying word form regions.

One of the advantages of the NeuroGen framework is its flexibility and capacity - one can provide an arbitrary target response map containing desired activation levels for any (or all) of the 24 brain regions that have encoding models and produce synthetic images that achieve that vector as closely as possible. As a simple example, we performed joint optimization of two or three regions, where we maximized and/or minimized their activations together. We chose to use V1v as one of the regions as this is known to activate in response to high-frequency patterns and results could be readily validated visually. Indeed we do see that when maximizing V1v and face/place areas, we get faces/places with an abundance of texture and when minimizing V1v we get places/faces with flat features. From looking at the scatter plots representing the synthetic and natural images' predicted activations in Figure \ref{fig:6}, there is a clear separation of the two, where the synthetic images clearly push the predicted brain activations to levels not achievable by the best-matching natural images. This example application of NeuroGen highlights another advantage of this framework in that one could synthesize stimuli predicted to evoke response patterns not generally observed in response to natural images. 

\subsection*{Limitations and Future Work}
There are a few limitations in this work. First, the range of the synthetic images is constrained by the images on which the generative network is trained, in this case ImageNet. Any preferences that exist for image content or features in the encoding model that do not exist in the ImageNet database may remain obscured in NeuroGen. Second, the deep generative network has a parameter that controls the balance between fidelity and variety of the synthetic images produced. It could be that varying this parameter would provide more realistic images, but it may also result in images that do not have as extreme predicted activation and/or have less variety and thus contain less information about the underlying stimuli-response landscape. Third, the optimization of the synthetic images is done in two steps, by first selecting the top 10 image classes and then optimizing the noise vector in that image class space. The classes identified in the first step could be constraining the synthesizer so that it is not identifying a global optimum; however, this trade-off was deemed an acceptable sacrifice for the more natural-looking images provided by a conditional generator. Lastly, this study employed AlexNet but more recent studies have found that other recent state-of-the-art methods like ResNet \cite{he2016deep} and VGG19 \cite{simonyan2014very} can perform better in terms of neural predictivity. Exploring these architectures can also be useful in subsequent studies. 

The NSD data on which the encoding model was trained is unsurpassed in its quality and quantity, consisting of densely-sampled fMRI in 8 individuals with several thousand image-response pairs per subject. Still, the natural images sourced from the COCO dataset used in the NSD experiments are inevitably limited in their content and features, which can mean possibly inaccurate brain-response mappings for images not used to train the encoding model. Additionally, when calculating preference ratios in the measured NSD data, it was at times difficult to choose the combination of image labels that produced the desired image content or features (e.g. only a person's face). Relatedly, it is not always straightforward to classify the natural or synthetic images into the appropriate category; the word form areas were particularly challenging. In addition, fMRI has many known sources of noise/confounds such as system-related instabilities, subject motion and possibly non-neuronal physiological effects from breathing and blood oxygenation patterns \cite{LIU2016141}. Careful design of the acquisition and post-processing pipeline for the NSD data mitigated these effects. Finally, the localizer task, while previously validated, may have some variability due to the contrast threshold applied. Using a more or less liberal threshold for the region boundary definition may result in different results than what is presented here. Different visual regions used in this work did have some overlap within certain individuals, which could have contributed to similarities in synthetic image content for regions of different categories. Supplementary Figure S14-22 show each individual's regional definitions and a heatmap of the Dice overlap of regions from different categories for each individual. 

To validate and demonstrate the capability of our novel NeuroGen framework, we present here as a proof-of-concept optimization of predicted responses in one, two or three regions. However, this optimization can be performed on an arbitrary desired activation pattern over any regions (or voxels) that have existing encoding models. Generative networks for creating synthetic images are an highly active area of research; specialized generators for faces or natural scenes could be integrated into the NeuroGen framework to further improve the range and fidelity of the synthetic images. Furthermore, the work presented here relies on predicted activation responses; no testing of the measured responses in humans to NeuroGen's synthetic images was examined. Future work will involve presentation of these synthetic images to individuals while undergoing fMRI to test if their responses are indeed more extreme than the best natural images. One hypothesis is that the synthetic images may command more attention, as it is clear they are not perfectly natural and thus may produce a more extreme response than natural images, as found in studies of single or neuronal population responses \cite{PONCE2019999, Bashivaneaav9436}. The other is that there may be some confusion about what the image contains or additional processing that an individual will undergo when interpreting the image that will result in an unpredictable pattern of response. If it can be demonstrated that synthetic images indeed produce activations matching a pre-selected target pattern, the NeuroGen framework could be used to perform macro-scale neuronal population control in humans. Such a novel, noninvasive neuromodulatory tool would not only be powerful in the hands of neuroscientists, but could also open up possible avenues for therapeutic applications.

\subsection*{Conclusions}
The NeuroGen framework presented here represents a robust and flexible framework that can synthesize images predicted to achieve a target pattern of regional activation responses in the human visual cortex that exceeds that of predicted responses to natural images. We posit that NeuroGen can be used for discovery neuroscience to uncover novel stimuli-response relationships. If it can be shown with future work that the synthetic images actually produce the desired target responses, this approach could be used to perform macro-scale, non-invasive neuronal population control in humans.

\section*{Material and Methods}
\subsection*{Natural Scenes Data Set}
We used the Natural Scenes Dataset (NSD; \url{http://naturalscenesdataset.org}\cite{Allen2021.02.22.432340} to train the encoding model. In short, the NSD dataset contains densely-sampled functional MRI (fMRI) data from 8 participants collected over approximately a year. Over the course of 30-40 MRI scans, each subject viewed 9,000–10,000 distinct color natural scenes (22,000–30,000 trials with repeats) while undergoing fMRI. Scanning was conducted at 7T using whole-brain gradient-echo EPI at 1.8-mm iso-voxel resolution and 1.6 s TR. Images were sourced from the Microsoft Common Objects in Context (COCO) database \cite{lin2014microsoft}, square cropped, and presented at a size of 8.4° $\times$ 8.4°. A set of 1,000 images were shared across all subjects; the remaining images for each individual were mutually exclusive across subjects. Images were presented for 3s on and 1s off. Subjects fixated centrally and performed a long-term continuous recognition task on the images in order to encourage maintenance of attention. The fMRI data were pre-processed by performing one temporal interpolation (to correct for slice time differences) and one spatial interpolation (to correct for head motion). A general linear model was then used to estimate single-trial beta weights, representing the voxel-wise activation in response to the image presented (beta version 3, using GLMdenoise and ridge-regression from NSD). Cortical surface reconstructions were generated using FreeSurfer, and both volume- and surface-based versions of the response maps were created.

In addition to viewing the COCO images, individuals all underwent the same image functional localizer (floc) task to define the visual region boundaries \cite{Stigliani12412}. In short, regions of interest were defined by contrasting activation maps for different types of localizer images (floc-bodies, floc-faces, floc-places, floc-words). Several regions exhibiting preference for the associated category were defined (e.g., floc-faces was based on t-values for the contrast of faces$>$non-faces). Regions were defined by drawing a polygon around a given patch of cortex and then restricting the region to vertices within the polygon that satisfy t$>$0. See Supplementary Figure S14-22 for the 8 individuals' early and late visual region maps, along with quantification of the regions' overlaps using Dice coefficient.

\subsection*{Deepnet feature-weighted receptive field encoding model}
We trained a Deepnet feature-weighted receptive (fwRF) encoding model\cite{st2018feature} using the paired NSD images and fMRI response maps described above. Here, instead of the voxel-wise model previously created, we trained a single model for each of the 24 early and late visual regions for each of the 8 individuals in the NSD dataset. There are three components in the Deepnet-fwRF model: $K$ feature maps, a vector of feature weights $w_{k}$, and a feature pooling field.The output of each fitted encoding model is the predicted activation $\hat{r}$ for a given region for a given individual in response to an image $S$:
$$
\hat{r}(S) = \sum_{k}^{K}w_{k}\int_{-D/2}^{D/2}\int_{-D/2}^{D/2} g(x,y;\mu_{x}, \mu_{y}, \sigma_{g})f_{i(x)j(y)}^{k}(S)dxdy
$$
where $w_{k}$ is the feature weight for $k$th feature map $f^k$, $g$ is the feature pooling field (described below), $D$ is the total visual angle sustained by the image, $i(x) = \lfloor (2x+D)/2\Delta \rfloor$ (likewise for $j(y)$) is the discretization depending on $\Delta=D/n_{k}$ which is the visual angle sustained by one pixel of a feature map with resolution, and $n_{k} \times n_{k}$ is the resolution of $k$th feature map.

The feature maps $f^k$ were obtained from Alexnet\cite{krizhevsky2012imagenet}, a deep convolutional neural network containing 5 convolutional layers (interleaved with max-pool layers) and 3 fully-connected layers. AlexNet was originally trained for classification of images in ImageNet\cite{russakovsky2015imagenet}, and is often used to extract salient features from images. The exact structure and trained network can be downloaded as part of the Pytorch library, and is also available at \url{https://github.com/pytorch/vision/blob/master/torchvision/models/alexnet.py}. The feature maps can be drawn from all convolutional layers and fully-connected layers in Alexnet. To limit the total number of feature maps, we first set the maximum feature maps for each layer to 512. For those layers whose dimension exceeded 512, we calculated the the variance of the layer values across the image set and retained those 512 feature maps with the highest variance. We then concatenated the selected feature maps having the same spatial resolution, which resulted in three feature maps of size $(256, 27, 27)$, $(896, 13, 13)$ and $(1536, 1, 1)$. 

The fwRF model was designed based on the hypothesis that the closer the feature map pixel is to the center of the voxel's feature pooling field, the more it contributes to the voxel's response. We also assume this is the case for clusters of voxels (regions in our case). The feature pooling field was modeled as an isotropic 2D Gaussian blob:
$$
g(x,y;\mu_{x}, \mu_{y}, \sigma_{g}) = \frac{1}{\sqrt{2\pi}\sigma_{g}}\exp \left[-\frac{(x-\mu_{x})^{2}+(y-\mu_{y})^{2}}{2\sigma_{g}^{2}} \right ]
$$
where $\mu_{x}$, $\mu_{y}$ are the feature pooling field center, and $\sigma_{g}$ is the feature pooling field radius. The feature pooling field center and radius were considered hyperparameters and learned during training of the encoding model. By definition, when the radius of the feature pooling field is very large (e.g. $\sigma_{g} \gg \Delta$), the predicted activation from a single layer reduces to a weighted sum of all pixels within the field; otherwise, it reduces to just one single spatial unit. In experiments, the grid of candidate receptive fields included 8 log-spaced receptive field sizes between 0.04 and 0.4 relative to $1/n_{k}$, and the candidate feature pooling field centers were spaced 1.4 degrees apart (regardless of size), resulting in a total of 875 candidate feature pooling fields. We searched the regularization parameter over 9 log-spaced values between $10^{3} \sim 10^{7}$ and chose the one that had the best performance on the held-out validation set.

Each of the 2688 image feature maps has an associated feature weight, $w_{k}$, indicating how important the $k$th feature map was for predicting that region's activity. Once the feature pooling hyperparameters were determined, a set of feature weights were then learned via ridge regression for each visual region in each of the 8 individuals. Since the images that each individual viewed were different and the individuals had a slightly different number of image-activation pairs, we used the individual-specific images for training and validating the ridge models and tested the ridge models on the shared 1000 images where we calculated the predicted accuracy. During training, we held 3,000 randomly selected image-activation pairs as a validation set to choose the best hyperparameters.

\subsection*{The BigGAN-deep image generator network}
We used the generator from BigGAN-deep, a pretrained deep generative adversarial network (GAN), utilizing the ImageNet image set, whose goal is to synthesize images of a given category that look natural enough to fool an automated fake/real classifier\cite{brock2018large}. The BigGAN-deep was built upon the self-attention GAN (SA-GAN)\cite{zhang2019self} with some differences: i) a shared class embedding was used for the conditional BatchNorm layers to provide class information, ii) the entire noise vector $z$ was concatenated with the conditional vector to allow the generator $G$ to use the latent space to directly influence features at different resolutions and levels of hierarchy, iii) the noise vector $z$ was truncated by resampling the values with magnitude above a chosen threshold to fall inside that range instead of using a Normal or Uniform distribution and iv) orthogonal regularization was used to enforce the models' amenability to truncation. BigGAN-deep's truncation threshold controls the balance of fidelity and variety of the synthetic images: larger thresholds lead to higher variety but lower fidelity images while lower thresholds lead to higher fidelity images of lower variety. In our experiments, the truncation parameter was set to 0.4 to achieve a balance of both fidelity and variety. The PyTorch version of the pretrained model can be publicly downloaded at \url{https://github.com/huggingface/pytorch-pretrained-BigGAN}.

\subsection*{Synthesizing images to optimize regional predicted activation}
We aimed to synthesize images using our generator network $G$ that either maximized the activation in single or pairs of regions or maximized activation in one region while minimizing activation in another region, depending on the experiment. For the sake of simplicity, we will describe the optimization procedure using the example goal of maximizing activation in a single region. Because our generator network (BigGAN-deep) is a conditional GAN, it requires identification of an image class via a one-hot encoded class vector $c$, then uses a noise vector $z$ to generate an image of that class. We performed the optimization in two steps; first, we identified the 10 most optimal classes, then, for each class we optimized over the noise vector space. To identify optimal classes, we generated 100 images from each of the 1000 classes in the ImageNet database using 100 different random noise vectors. We input the resulting images into the encoding model to obtain associated predicted regional activation, which was averaged over the 100 images per class. The 10 classes that gave the highest average predicted activation for the region of interest were identified and encoded in $c_i$ ($i = 1,...10$), and corresponding image generator noise vectors $z_{ij}$ ($i = 1,...10$, $j = 1,...10$) were further optimized via backpropagation with 10 random initialization seeds per class. The result of this optimization is a set of 100 images $G(c_i,z_{ij})$ that yielded maximal predicted activation of the target region. Formally, and including a regularization term, we posed the optimization problem as finding the codes $\hat{z}_{ij}$ such that:
$$
\hat{z}_{ij}(c_i) = \arg\max_{z_{ij}}(\hat{r}_{t}(G(c_i,z_{ij})) - \lambda \Vert z_{ij} \Vert )
$$
where $\lambda=0.001$ was the regularization parameter. For the maximization/minimization of region pairs, the cost function aimed to maximize/minimize the sum of the two regions' activations together, with both regions considered equally. For the maximization of one region and the minimization of the other, the cost function was the sum of the max and min, with both regions considered equally. Once this optimization was performed, we selected the top 10 images by taking the most optimal image from each of the 10 distinct classes.

\section*{Data availability}
Information on the NSD dataset is available at \url{http://naturalscenesdataset.org}. The dataset will be made publicly available upon the NSD manuscript publication.

\section*{Code availability}
Code is available at \url{https://github.com/zijin-gu/NeuroGen}.

\bibliography{scifile}

\bibliographystyle{naturemag}

\section*{Acknowledgements}
This work was funded by the following grants: R01 NS102646 (AK), RF1 MH123232 (AK), R21 NS104634 (AK), R01 LM012719 (MS), R01 AG053949 (MS), NSF CAREER 1748377 (MS), and NSF NeuroNex Grant 1707312 (MS). The NSD data were collected under the NSF CRCNS grants IIS-1822683 (KK) and IIS-1822929 (TN).

\section*{Author contributions statement}
A.K. and M.S. devised the project, the main conceptual ideas and proof outline. Z.G. worked out almost all of the technical details, and performed the analysis for the suggested experiment. K.J. and M.K. gave suggestions on data usage and experimental design. E.A., Y.W. and K.K. provided the data. T.N. provided the encoding model architecture. Z.G. and A.K. wrote the manuscript. All authors provided critical feedback and helped shape the research, analysis and manuscript.

\section*{Competing interests}
The authors declare no competing interests.

\section*{Citation gender diversity statement}
Recent work in several fields of science has identified a bias in citation practices such that papers from women and other minorities are under-cited relative to the number of such papers in the field \cite{dworkin2020extent}. Here we sought to proactively consider choosing references that reflect the diversity of the field in thought, form of contribution, gender, and other factors. We obtained predicted gender of the first and last author of each reference by using databases that store the probability of a name being carried by a woman \cite{dworkin2020extent}. By this measure (and excluding self-citations to the first and last authors of our current paper), our references contain $4.43\%$ woman(first)/woman(last), $25.48\%$ man/woman, $10.89\%$ woman/man, and $59.2\%$ man/man. This method is limited in that a) names, pronouns, and social media profiles used to construct the databases may not, in every case, be indicative of gender identity and b) it cannot account for intersex, non-binary, or transgender people. We look forward to future work that could help us to better understand how to support equitable practices in science.

\end{document}